\documentclass[reprint,amsmath,amssymb,apr,aip,onecolumn, 10pt]{revtex4-1}
\usepackage{graphicx}
\usepackage{graphics}
\usepackage{mathptmx}
\usepackage{times}
\usepackage{amsmath}
\usepackage{amssymb}
\usepackage{dcolumn}
\usepackage{bm}
\usepackage{units}
\usepackage{color}
\usepackage{soul}
\usepackage{xcolor}

\begin{document}
\title{Supervised Learning in Spiking Neural Networks with Phase-Change Memory Synapses}
\author{S. R. Nandakumar}  \affiliation{New Jersey Institute of Technology, Newark, NJ 07102, USA}\affiliation{IBM Research -- Zurich, 8803 R\"{u}schlikon, Switzerland.}
\author{Irem Boybat} \affiliation{IBM Research -- Zurich, 8803 R\"{u}schlikon, Switzerland.}
\affiliation{{Ecole Polytechnique Federale de Lausanne (EPFL), 1015 Lausanne,} Switzerland}
\author{Manuel Le Gallo} \affiliation{IBM Research -- Zurich, 8803 R\"{u}schlikon, Switzerland.}
\author{Evangelos Eleftheriou} \affiliation{IBM Research -- Zurich, 8803 R\"{u}schlikon, Switzerland.}
\author{Abu Sebastian}\email{ase@zurich.ibm.com} \affiliation{IBM Research -- Zurich, 8803 R\"{u}schlikon, Switzerland.}
\author{Bipin Rajendran}\email{bipin@njit.edu} \affiliation{New Jersey Institute of Technology, Newark, NJ 07102, USA}
\date{\today}

\begin{abstract}
Spiking neural networks (SNN) are artificial computational models that have been inspired by the brain's ability to naturally encode and process information in the time domain. The added temporal dimension is believed to render them more computationally efficient than the conventional artificial neural networks, though their full computational capabilities are yet to be explored. Recently, computational memory architectures based on non-volatile memory crossbar arrays have shown great promise to implement parallel computations in artificial and spiking neural networks. In this work, we experimentally demonstrate for the first time, the feasibility to realize high-performance event-driven in-situ supervised learning systems using nanoscale and stochastic phase-change synapses. Our SNN is trained to recognize audio signals of alphabets encoded using spikes in the time domain and to generate spike trains at precise time instances to represent the pixel intensities of their corresponding images. Moreover, with a statistical model capturing the experimental behavior of the devices, we investigate architectural and systems-level solutions for improving the training and inference performance of our computational memory-based system. Combining the computational potential of supervised SNNs with the parallel compute power of computational memory, the work paves the way for next-generation of efficient brain-inspired systems.
\end{abstract}

\maketitle In recent years, deep learning algorithms have become successful in solving complex cognitive tasks surpassing the performance achievable
by traditional algorithmic approaches, and in some cases, even expert humans. However, conventional computing architectures are confronted with
several challenges while implementing the multi-layered artificial neural networks (ANNs) used in these algorithms,  especially when compared against
the approximately $20\,$W power budget of the human brain. The inefficiencies in the von Neumann architecture for neural network implementation arise
from the high-precision digital representation of the network parameters, constant shuttling of {large amounts of} data between processor and memory,
and the ensuing limited computational parallelism and scalability. In contrast, the human brain employs billions of neurons that communicate with
each other in a parallel fashion, through dedicated, analog, and low-precision synaptic connections. The spike-based data encoding schemes used in
these biological networks render the computation and communication asynchronous, event-driven, and sparse, contributing to the high computational
efficiency of the brain.

The size and complexity of artificial neural networks are expected to continue to grow in the future and has motivated the search for efficient and
scalable hadware implementation schemes for learning systems\cite{Lecun2015}. Spiking neural networks (SNNs) are excellent candidates to implement
large learning networks, especially for energy and memory-constrained embedded applications, as they closely mimic some of the key computational
principles of the brain. Application specific integrated circuit (ASIC) designs such as TrueNorth from IBM\cite{Y2014merollaScience} and Loihi from
Intel\cite{Y2018DaviesLoihi}, that implement SNN dynamics, have been successful in demonstrating two to three orders of magnitude energy efficiency
gain by mimicking the sparse, asynchronous, and event-driven nature of computation in the brain. However, the area expensive static random access
memory (SRAM) circuits used for synaptic weight storage in these chips limit the amount of memory that can be integrated on-chip and hence the
scalability of these architectures.

\begin{figure}[h!]
\centering
\includegraphics[width = 1.0\textwidth]{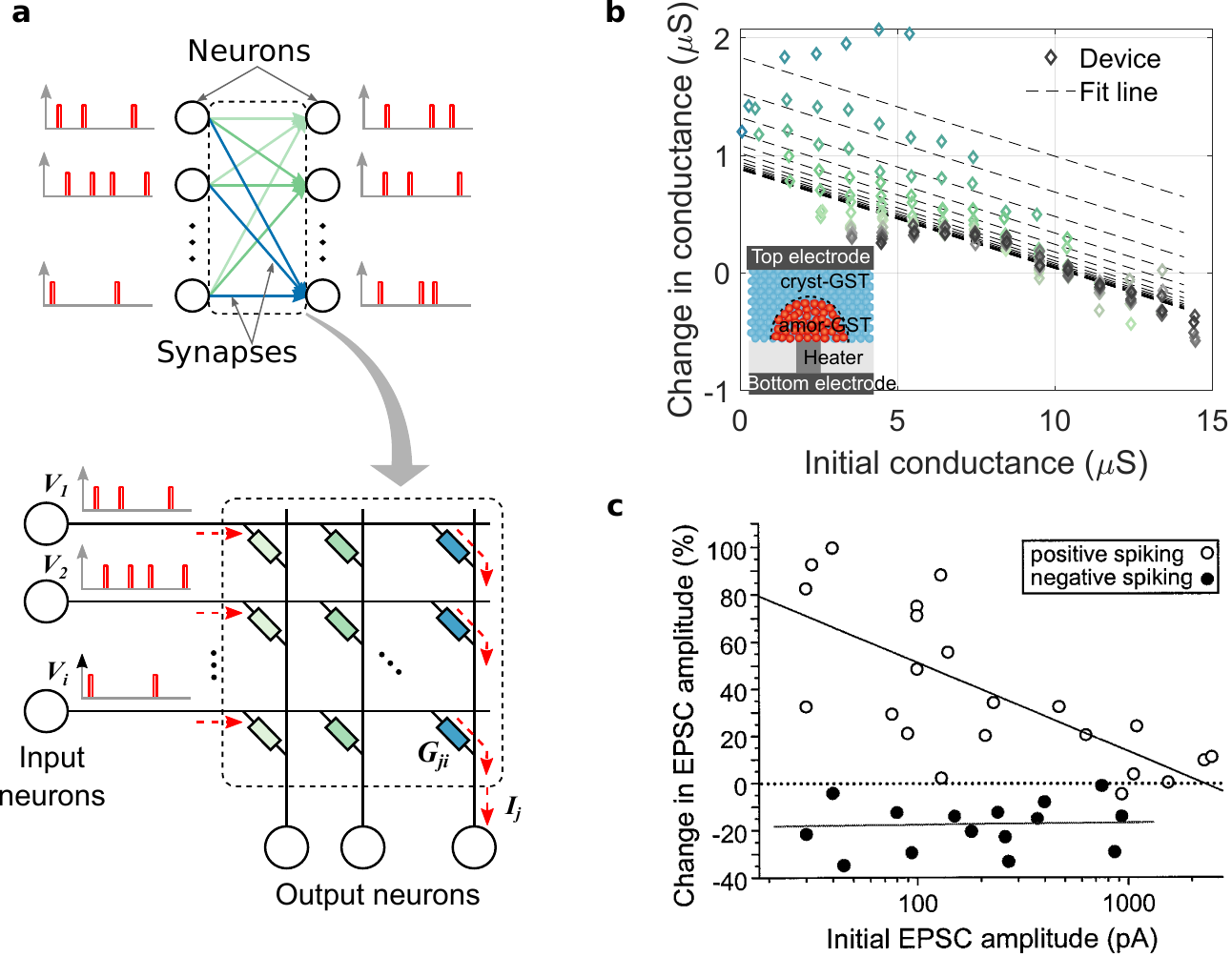}
\caption{\textbf{Spiking neural network implementation using computational memory array.} \textbf{a} A single layer spiking neural network that
translates an input set of spike trains to an output set of spike trains (top). The network connectivity matrix could be realized using a
non-volatile computational memory array (bottom). The voltage spike trains $V_{i}$ are applied along the word-lines and the weighed summations are
read as currents $I_j$ from the bit-lines. \textbf{b} The characteristic state-dependent behavior of average conductance change observed in a
phase-change memory (PCM) device. The device structure in the inset illustrates the amorphous region (amor-GST) formed inside the crystalline region
(cryst-GST). \textbf{c} The synaptic conductance changes measured using changes in excitatory postsynaptic current (EPSC) as a function of its
initial EPSC amplitudes, from the hippocampal neurons in a rat\cite{Bi1998}. The state-dependent nature of conductance change in response to positive
(causal) spiking is analogous to that observed in the PCM devices.} \label{fig:intro_PCM}
\end{figure}

Crossbar arrays of analog non-volatile memory devices can perform weighted summation of its word line voltages in parallel using the device
conductances and the results are available as currents at its bit lines. This memory architecture performing computations (computational memory) is
using a combination of  Ohm's law and Kirchhoff's law to reduce matrix-vector multiplications to $\mathcal{O}(1)$ complex operations
\cite{Y2017burrAPX, Y2018legalloNatureElectronics, Y2018sebastianJAP,Y2019xiaNatureMaterials}. Neural networks have layers of neurons each of which
receive a weighted summation of neuronal response from its previous layer. The underlying   matrix-vector multiplications are   computationally
expensive  in traditional digital systems due to their large sizes and the necessity to store these matrices off chip. SNNs processing asynchronous
events in time can significantly benefit from an on-chip computational memory that could store the synaptic weights in the device conductance values
and provide dedicated connectivity patterns to process parallel events in real-time (Fig.~\ref{fig:intro_PCM}\textbf{a}). For instance, such
computational memory based SNNs could also be directly interfaced with spike encoding sensors such as artificial retina\cite{Lichtsteiner2008DVS} or
cochlea\cite{SCLiu2014} to process asynchronous binary spike streams of real-world signals.

However, achieving software-equivalent performance using computational memory realized using today's memory devices is challenging, due to the
inherent non-idealities in the conductance modulation characteristics of these nanoscale devices.  Phase-change memory (PCM) is a mature non-volatile
memory technology that has demonstrated gradual conductance modulation, however it exhibits several non-ideal characteristics that are typical to
most nanoscale memories, including limited precision, stochasticity, non-linearity, as well as drift of the programmed conductance states with time
\cite{Y2015burrITED, Y2018boybatNatComm}. Nonetheless, large-scale experiments demonstrating the effective use of PCM as synapses in ANNs show
significant promise \cite{Y2015burrITED, Y2018ambrogioNature}. PCM is an attractive technology also for SNN implementations
\cite{Y2011kuzumNanoletters, Y2013jacksonACM, Y2016tumaEDL, Y2017sidlerICANN, Y2018boybatNatComm} and the similarity of the state-dependent nature of
the conductance update in PCM and in a biological synapse (Fig.~\ref{fig:intro_PCM}\textbf{b}, {c}) opens up the possibility of exploiting the device
physics rather than merely being limited by them. Most of the research efforts and experimental demonstrations using PCM in SNNs focus on an
unsupervised training based on a local learning rule observed in biology known as spike-timing-dependent plasticity (STDP)\cite{Bi1998}. However,
unsupervised STDP based learning generally yields sub-par results in comparison to supervised training or has been limited to problems where the
desired response of the neural network is not known beforehand\cite{Diehl2015}. There is also a growing body of evidence from neuroscience literature
suggesting that data encoding using precise spike times in biological neural networks\cite{Mainen1995a, Y1997Reich, y2004Uzzell} have several
computational advantages compared to rate based encoding schemes\cite{Maass1997}.

In this article, we focus on in-situ supervised training of SNNs that learn to generate spikes that encode data corresponding to real-world signals
using precise spike times and experimentally demonstrate their hardware implementation using more than 177,000 on-chip PCM devices. Moreover, we
capture the statistical behavior of PCM devices with accurate models and use them to evaluate the improvement in training performance as a function
of the number of PCM devices used per synapse. Next, we examine how modifications to the input encoding scheme with random jitter can improve
learning and lastly, we demonstrate an array-level compensation scheme to tackle the accuracy drop due to temporal evolution of PCM-based synapses.

\section*{Results}
\subsection{SNN learning experiment}\label{expt:implementation}

\begin{figure}[h!]
\centering
\includegraphics[width = 1.0\textwidth]{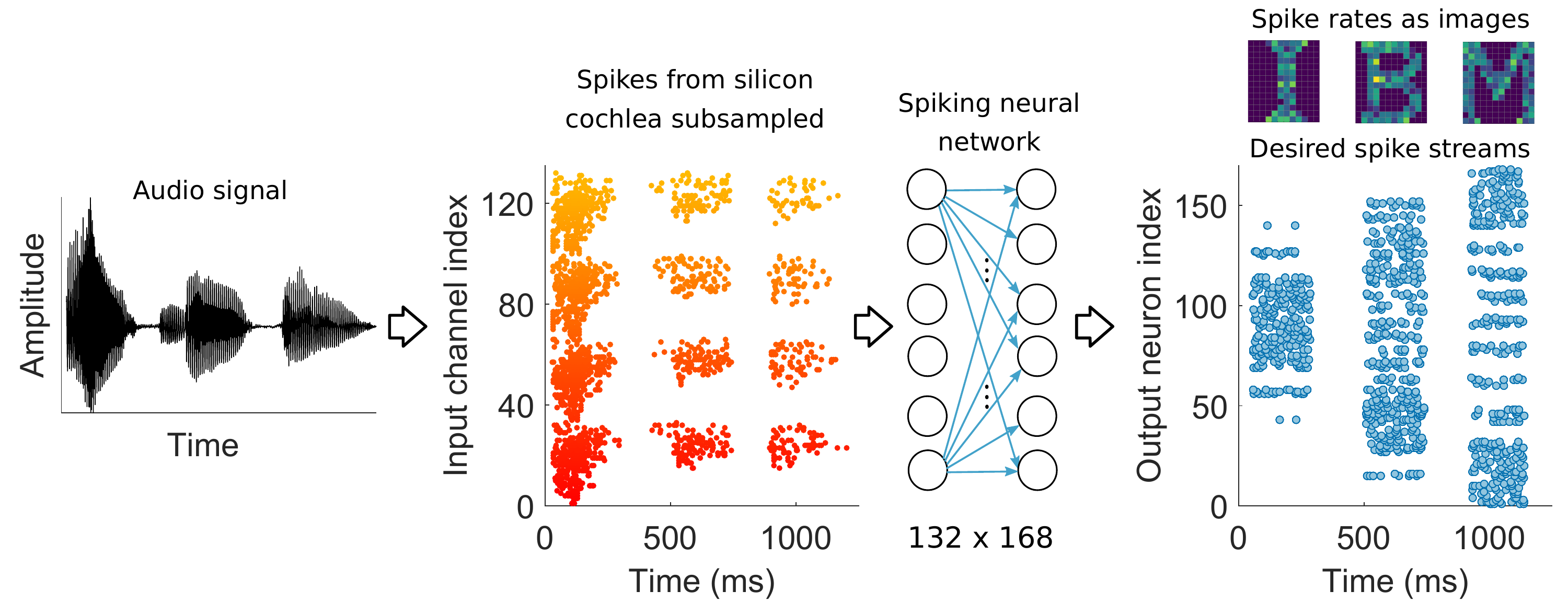}
\caption{\textbf{SNN training problem.} The audio signal is passed through a silicon cochlea chip to generate spike streams. These spike streams are
sub-sampled and applied as input to train the single layer SNN. The desired spike response from the networks representing the images ($14\times12$
pixels) corresponding to the characters in the audio is also shown.} \label{fig:SNN_train_net}
\end{figure}

The training problem and the network we used for the experiment are illustrated in Fig.~\ref{fig:SNN_train_net}. The learning task of the network is
to recognize and translate audio signals corresponding to spoken alphabets into corresponding images, with all information encoded in the spike
domain, as described below.   An audio signal captured when a human speaker  utters the characters `IBM' (Eye..Bee..Em) is converted to a set of
spike streams using a Silicon cochlea chip\cite{SCLiu2014}  and the resulting 132 spike streams (representing the signal components in 64 frequency
bands) are subsampled to an average spike rate of 10\,Hz  to generate the binary spike inputs to the network (see Methods for more details). A raster
plot of the generated spikes is shown in Fig. \ref{fig:SNN_train_net}. At the output of the network, there are 168 spiking neurons, with the spike in
each neuron representing the instantaneous pixel intensity of the image corresponding to the input audio signal. The desired spike stream from each
output neuron is obtained from a Poisson random process whose arrival rate is chosen to be proportional to the corresponding pixel intensities in the
images (14$ \times $12 pixels showing the characters `I', `B', and `M'), inspired by similar statistical distributions  observed in animal
retina\cite{y2004Uzzell}. Each image has an average duration of 230\,ms and is mapped to the corresponding time window in the audio signal. The
network hence receives 132 spike streams corresponding to the audio signals and is connected to 168 spiking neurons at the output, corresponding to
the pixels of the image. In the experiment, the synaptic strength between the input streams and the output neurons is represented using the
conductance of the PCM devices.

An input spike, arriving at time $t_i$  on an input synapse, triggers a  current flow into the output neuron. The synaptic current in response to
each spike is modeled as $ I_{ker}(t) =  (e^{-(t-t_i)/\tau_1} - e^{-(t-t_i)/\tau_2})u(t-t_i) $ multiplied by synaptic weight $ W $, where $u(t)$ is the Heaviside step function
(with $ \tau_1 = 5\,$ms  and $ \tau_2 = 1.25\,$ms).  The sum of all the weighted currents are integrated by leaky-integrate and fire (LIF) neurons to
determine a voltage analogous to the membrane potential of the biological neurons. When this voltage exceeds a threshold, it is reset to a resting
potential and a spike is assumed to be generated. During the course of training, PCM conductance values read from hardware are used to calculate the
synaptic currents  and the neuronal dynamics are implemented in software. A supervised training algorithm is used to determine the desired weight
updates such that the observed spikes from the SNN are at the desired time instances. The weight updates will be implemented by modulating the
corresponding PCM conductance values by applying a sequence of programming pulses. We avoid verifying if the observed conductance change matches the
desired update. This blind programming scheme (without expensive read-verify) is expected to be the norm of computational memory based learning
systems in the future and in this study we experimentally evaluate the potential of analog PCM conductance to precisely encode spike time information
in SNNs.

\subsection{Phase-change memory synapse}
For our on-chip training experiment, we used a prototype chip containing more than one million doped-Ge$_2$Sb$_2$Te$_5$ (GST) based PCM devices
fabricated in 90\,nm CMOS technology node\cite{Y2010closeIEDM}. The GST dielectric has a lower resistivity in its poly-crystalline state and a high
resistivity in its amorphous phase. An amorphous region is created around the narrow bottom electrode via a melt-quench process. Its conductance can
be gradually increased by a sequence of partial-SET pulses applied to the device. A threshold switching phenomenon permits large current flowing
though the amorphous volume to increase its temperature and to initiate crystal growth. We have characterized the crystal growth driven conductance
evolution in the PCM array and have created statistically accurate models\cite{Y2018nandakumarJAP}. The PCM models are used to pre-validate the experiment and to evaluate methods to improve training performance.

While the conductance increment (SET) operation in PCM can be gradual and accumulative, the melt-quench driven conductance decrement (RESET) process
is  non-accumulative. This leads to an asymmetric update behavior in conductance increase and decrease, necessitating the use of the standard
differential configuration for weight updates\cite{Suri11}. In this scheme, each network weight $ W $ is realized as the difference of two PCM
conductance values $ G_p $ and $ G_n $ ($ W = \beta (G_p - G_n) $ where $ \beta $ is a scaling factor to be implemented in the peripheral circuit of
the computational memory array).  This allows both increment and decrement of the $ W $ to be implemented as partial-SET operations on $ G_p $ and $
G_n $, respectively. This differential configuration improves the symmetry of weight updates and partially compensates the conductance
drift\cite{Suri2013NanoArch}. Further improvement in conductance change granularity, stochasticity, and drift behavior can
be achieved via a multi-PCM configuration\cite{Y2018boybatNatComm, Y2018BoybatNVMTS}.  In our training experiment, both the $ G_p $ and $ G_n $ are
realized as the sum of four PCM devices. For each synaptic update desired by the training algorithm, only one of the four devices is programmed,
chosen cyclically so that on average all devices receive   approximately equal number of update pulses~\cite{Y2018boybatNatComm}. The energy overhead
from the multiple devices per synapse is not expected to be significant since PCM devices can be read with low energy (1 -- 100 fJ per device)
\cite{Y2017legalloIEDM} and only one of the devices is programmed per update as in a conventional synapse. Although we are increasing the area for
each synapse, it is worth noting that typical computational memory based design area for neural networks are dominated by the circuits for peripheral
neurons rather than the synapse. Moreover, PCM devices have been shown to scale to nanoscale dimensions \cite{Y2011xiongScience} and through
technology scaling, the synaptic area could reduce significantly \cite{Y2012choiISSCC}. Thus in our implementation, each synapse is realized using 8
PCM devices, making a total of 177,408 devices to represent the weights of 22,176 synapses in the network.

\subsection{Training algorithm}
The supervised training of SNNs is a challenging task as the gradient descent based backpropagation algorithms do not apply directly due to the
non-differentiable dynamical behavior of spiking neurons (i.e., the membrane potential encounters a discontinuity at the point of spike). One
approach to circumvent this limitation is to train a continuous-valued ANN using standard backpropagation algorithm and then convert it into a
SNN\cite{Cao2015, Diehl2016, Rueckauer2017}. However, in this method, the input data and neuron activations in the ANN are translated to spike rates
in the SNN, losing the advantage of precise time-based signal encoding, and necessitating longer processing times leading to sub-par performance and
energy efficiency\cite{Y2018PfeilFrontiers}. Also, the unconstrained training of the floating point synapses without taking into account the
non-idealities of analog memory devices will lead to further loss in accuracy when the the trained weights are transferred  to nanoscale synapses in
hardware. Moreover, training approaches that implement back-propagation in SNNs using approximate derivatives of the membrane potential around the
time of spikes are also aimed at minimizing cost functions, which have been described in terms of the output spike rate rather than precise spike
times\cite{Y2016_Pfeiffer, Wozniak2018}. Encoding events using precise spike times could be more efficient as it leads to sparse computations and low
latencies for decision making\cite{Bohte2002, Crotty2005, Gutig2006, Wang2016, Y2014merollaScience}.

Recently, several approximate spike time based supervised training algorithms have been proposed of varying computational complexity that have
demonstrated various degrees of success in benchmark problems in machine learning. {Among these, SpikeProp\cite{Bohte2002} is designed to generate
single spikes, Tempotron\cite{Gutig2006} uses a non-event driven error computation, and ReSuMe\cite{y2010Ponulak} and NormAD\cite{Y2015anwaniIJCNN}
(with relatively higher convergence rate) are designed to generate spikes at precise time instances via spike driven weight updates.}  In our
experiment, we use the {normalized approximate descent} (NormAD) algorithm which has been successful in achieving high classification accuracy for
the MNIST hand-written digit recognition problem\cite{Kulkarni2018}. According to this algorithm, the weight updates $ \Delta W $ are computed in an
event-driven manner, using the relation

\begin{align}
\Delta W  = \eta \int_{0}^{T} e(t) \frac{\hat{d} (t)}{\Vert \hat{d}(t)\Vert} dt
\end{align}
where $ \eta $ is the learning rate, $ T $ is the pattern duration, and $ e(t) $ is the difference between desired and observed binary spike trains.
$ \hat{d} (t) $ is obtained by convolving the input spike stream $ S_i(t) $ with {$ I_{ker}(t) $} and an approximate impulse response of the LIF
neuron (see Methods). The weight updates are computed only at the time instants corresponding to a spike generated by the learning network, or the
instances where a spike was desired (i.e., when $e(t)\neq0$). These are accumulated over the training pattern duration (one epoch) and is used to
modulate the network weights. The $ \Delta W $s were converted to desired conductance changes using the scaling factor $ \beta $. The desired
conductance changes lying in the interval [0.1,~1.5]\,$ \mu $S  were mapped to amplitudes of 50\,ns programming current pulses from 40\,$ \mu $A to
130\,$ \mu $A. The smaller conductance changes were neglected. The conductance updates during the training were performed by blindly applying the
programming pulses without verifying if the observed conductance change matches the desired update.

\subsection{Training performance}\label{Evaluation}

\begin{figure}[h!]
\centering
\includegraphics[width = 1.0\textwidth]{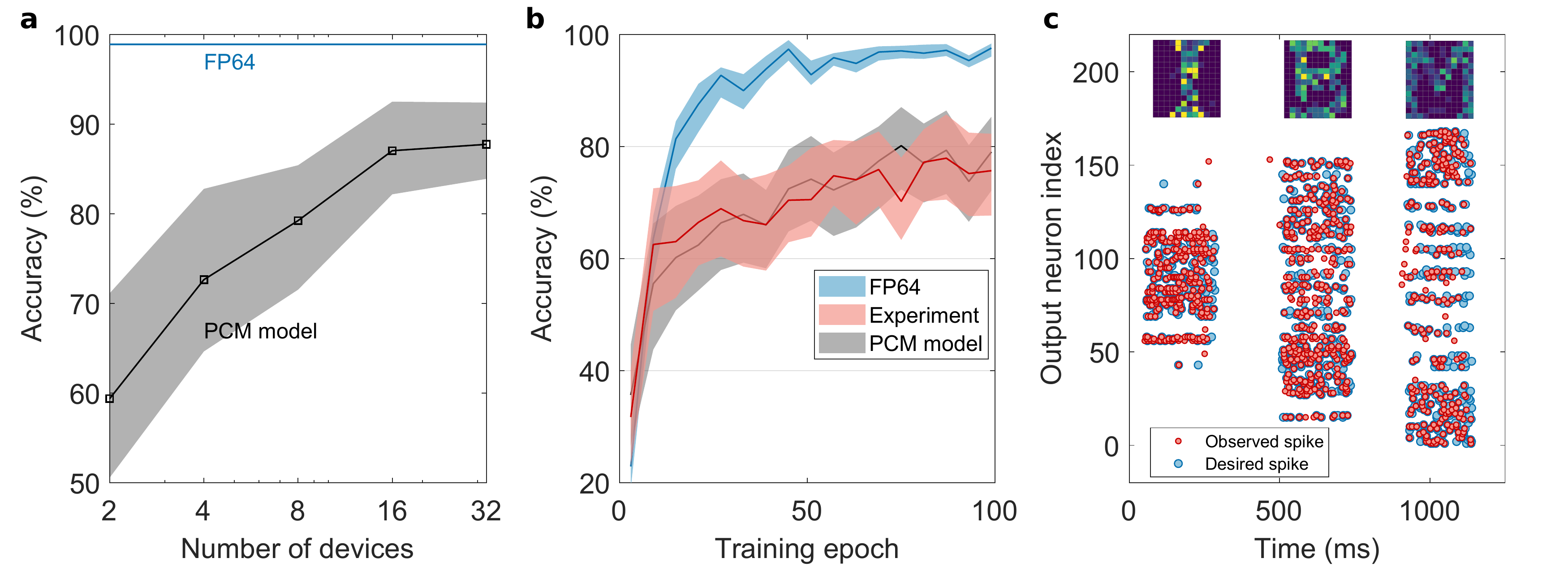}
\caption{\textbf{Training experiment using PCM devices.} \textbf{a} Simulated training accuracy as a function of the number of devices in a multi-PCM
synapse (92.5\% maximum accuracy). Accuracy is defined as the fraction of the spike events in the desired pattern corresponding to which a spike was
generated from respective output neurons within a certain time interval. The lower bound of the shaded lines correspond to 5\,ms interval, the middle
line to  10\,ms and the upper bound to 25\,ms.  \textbf{b} Accuracy as a function of training epochs from the experiment using on-chip PCM  devices.
Each synapse was realized using 8 PCM devices in differential configuration. The corresponding training simulation using the PCM model shows
excellent agreement with the experimental result. The experiment, PCM model, and the reference floating point (FP64) training achieve maximum
accuracies of 85.7\%, 87\%, and 98.9\% respectively for the 25\,ms error tolerance. \textbf{c} The raster plot of the desired and observed spike
trains from the trained network. A visualization of the character images whose pixel intensities are  generated from the observed spike rates is also
shown above the raster plot.} \label{fig:SNN_train_exp_result}
\end{figure}

First, we used the PCM model to pre-validate and optimize the training scheme. Fig.~\ref{fig:SNN_train_exp_result}\textbf{a} shows the improvement in
network training accuracy as the number of PCM devices used per synapse increases (in differential configuration). The performance of the network is
evaluated using an accuracy metric defined as the percentage of the number of spikes out of a total of 987 in the desired pattern which have an
observed spike from the SNN within a certain time interval. In the line plot of accuracy with shaded bounds, the lower bound, middle line, and the
upper bound respectively correspond to spike time tolerance intervals of 5\,ms, 10\,ms and 25\,ms. Note that the average output spike rate for each
of the character duration was less than 20\,Hz corresponding to an inter-arrival time of 50\,ms, and the task of the network is to create spikes each
one of which can be unambiguously associated with one of the target spikes. A fixed weight range obtained from the reference high-precision training
was mapped to the sum of conductance of 1 to 16 differential pairs and networks were trained for 100 epochs. Using more number of devices in
parallel, with only one programmed at each weight update, permitted smaller weight updates to be programmed more reliably.    Although the accuracy
was found to improve with  more PCM devices,  increasing the total number of devices beyond 16 in this problem did not lead to corresponding
improvements in accuracy. {One possible explanation is that, with more number of devices the observed conductance change (which has a limited dynamic
range for a chosen partial-SET programming scheme) captures smaller desired weight changes but neglects the larger desired weight changes, leading to
slower convergence.}  The maximum accuracy observed from the simulation was 92.5\% at 25\,ms timing error for 16 devices per synapse.

We performed the training experiment with the synapses realized using eight on-chip PCM devices in differential configuration and the SNN generated
more than 85\% of the spikes within the 25\,ms error tolerance (Fig.~\ref{fig:SNN_train_exp_result}\textbf{b}).  The training experimental results
agree well with the observations from the PCM model based simulation. The training accuracy obtained from the
corresponding 64-bit floating point (FP64) training simulation is also shown for reference. A raster plot of the spikes observed from the SNN trained
in the experiment is shown in Fig.~\ref{fig:SNN_train_exp_result}\textbf{c} as a function of time along with the desired spikes. The character images
shown on top are created using the average spike rate for the duration of each character and it indicates that the network was successfully trained
to generate the spikes to create the images.

\begin{figure}[h!]
\centering
\includegraphics[width = 1.0\textwidth]{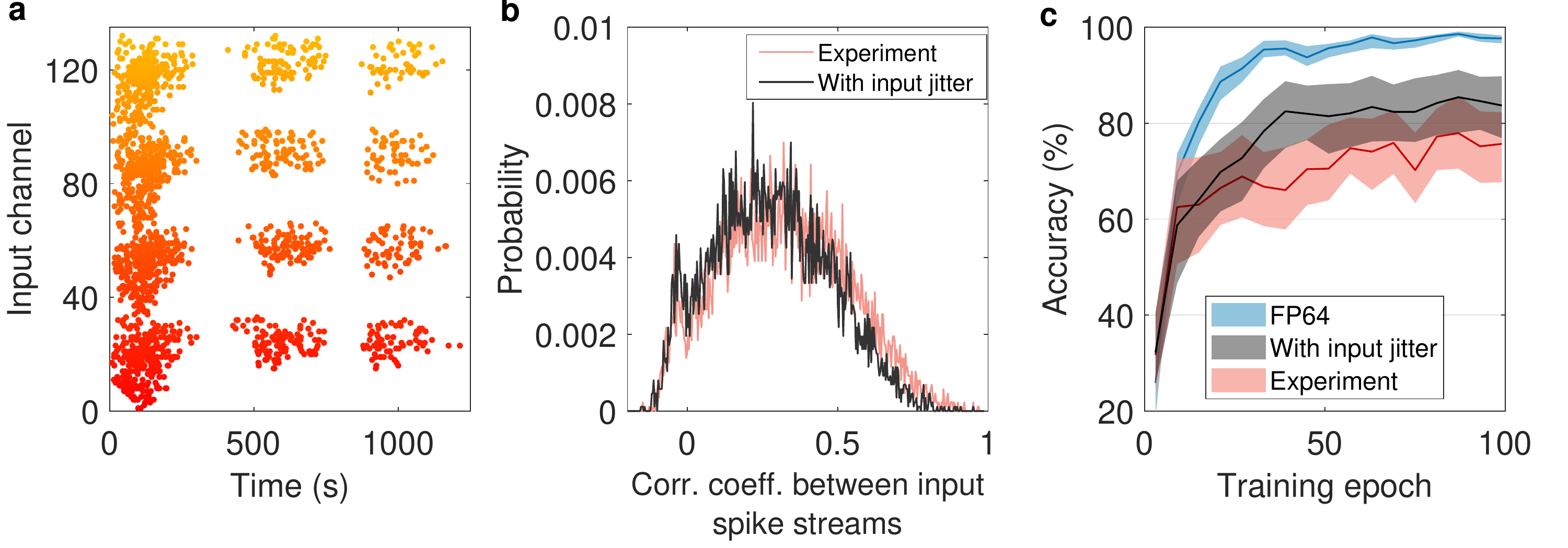}
\caption{\textbf{Role  of input correlations in network performance.} \textbf{a} Input spike streams with spike times jittered by random amounts
uniformly distributed in {[-25, 25]\,ms}. \textbf{b} The cross- correlation between the jittered spike streams are shifted towards zero compared to
the experimental input \textbf{c} The simulated training accuracy is improved when trained with input spike streams of reduced correlation.}
\label{fig:input_correlation}
\end{figure}

While the maximum accuracy obtained by the training the PCM devices is limited by the non-linearity, stochasticity, and granularity of its
conductance change, we observed that accuracy of the SNN could be further enhanced by modifying the input encoding scheme. The ability of a neural
network to classify its inputs depends on the correlation between the inputs.  In Fig.~\ref{fig:input_correlation} we show using the PCM model
simulation that the accuracy gap between those from the experiment and floating point training simulation can be reduced by decreasing the
correlation between the input spike streams. We added a random temporal jitter uniformly distributed in the interval [-25, 25]\,ms to each input
spike which causes the cross-correlation between the input spike streams to decrease. The correlation coefficients between the binary spike streams
were determined after smoothing them using a Gaussian kernel ($ e^{-t^2/2\sigma^2} $) of $ \sigma = 5\, $ms. Even though the added jitter only
reduces the correlation by a very small amount (Fig.~\ref{fig:input_correlation}\textbf{b}),  the training performance improves substantially,
suggesting that encoding schemes or network structures that inherently separate input features will improve training performance using low-precision
devices such as PCM.

\subsection{On-chip inference}
\begin{figure}[h!]
\centering
\includegraphics[width = 1.0\textwidth]{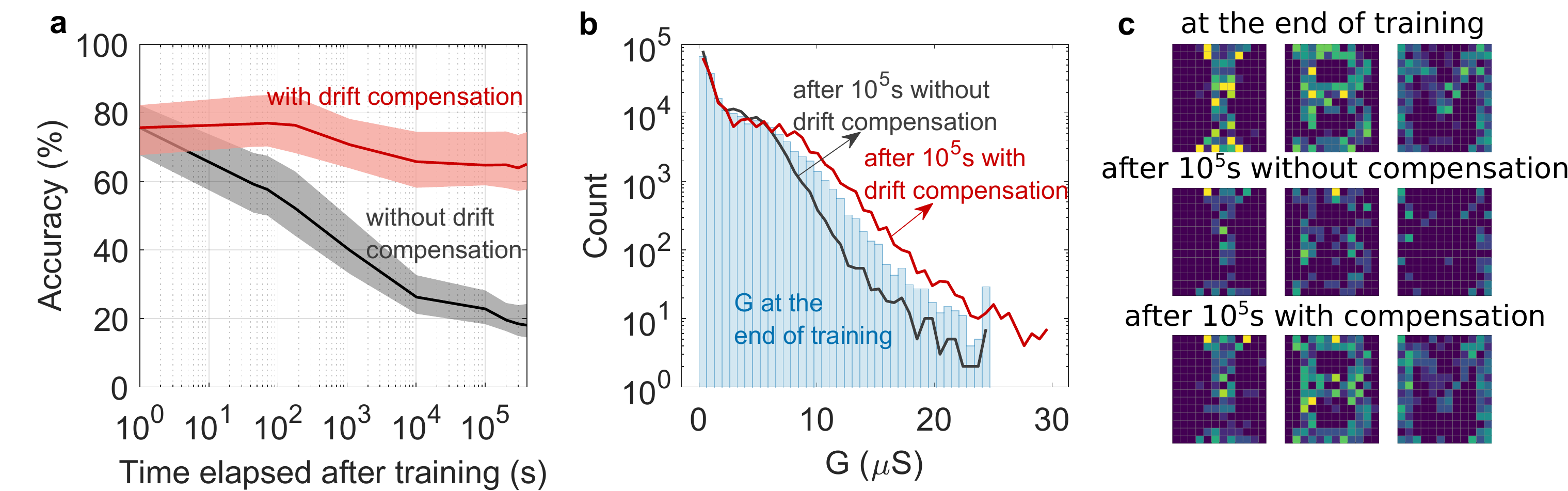}
\caption{\textbf{On-chip inference and drift compensation.} \textbf{a} Inference using trained PCM array. Due to conductance drift, the accuracy
drops over time (black line). The effect of drift can be compensated by a time-aware scaling method (red line). Percentage accuracy drop over 4$
\times 10^5 \,$s was reduced from 70\% to 13.6\% at 25\,ms error tolerance. \textbf{b} The drifted conductance distribution at the end of 10$ ^{5} $s
is compared with the trained conductance distribution. The effect of scaling on the drifted conductance is also shown.  \textbf{c} The images
generated by the SNN at the end of training for the audio input (top). The images generated after 10$ ^{5} $\,s (middle). The images generated with
drift compensation (bottom). The brightness of each pixel represents the spike rate for the duration of each character.} \label{fig:SNN_inf}
\end{figure}

The ability of a PCM based SNN to retain the trained state is evaluated by reading the conductance at logarithmic intervals of time and using it to
calculate the network response. Both the spike-time accuracy (Fig.~\ref{fig:SNN_inf}\textbf{a}) and the average spike rate (depicted as pixel
intensities in Fig.~\ref{fig:SNN_inf}\textbf{c}) drops due to conductance drift  over time (Fig.~\ref{fig:SNN_inf}\textbf{b}). The conductance
decrease causes the net current flowing into the neurons to reduce which result in errors in spike times and a drop in the neuron spike rate.
However, we show that this can be compensated via an array level scaling method described below.

The conductance drift in PCM is modeled using the empirical relation\cite{Karpov2007,Y2018legalloAEM}:
\begin{align}
G(t) & = G(t_0)\Big(\frac{t-t_p}{t_0-t_p}\Big)^{-\nu} \label{eq:drift_tp}
\end{align}
where $G(t)$ is the conductance of the device at time  $ t>t_0 $,  $ t_{p} $ denotes the time when it  received a programming pulse and $ t_{0} $
represent the time  instant at which its conductance was last read after programming.  Thus, each programming pulse effectively re-initializes the
conductance drift\cite{Y2018BoybatNVMTS}. As a result, the devices in the array will drift by different amounts during training, based on the instant
they received the last weight update. However, once sufficient time has elapsed after training, (i.e., when $ t $ becomes much larger than all the $
t_p $ values of the devices in the array),  the conductance drift can be compensated by an array level scaling. In our study, all the measured
conductances were scaled by  $ t_e^{0.035} $ where the $ t_e $ is the time elapsed since training and $ 0.035 $ is the effective drift coefficient
for the conductance range of the devices in the array. Fig.~\ref{fig:SNN_inf} shows the improvement in spike-time accuracy and spike rate obtained
using this scaling method. {The drop in accuracy after the compensation can be attributed to the conductance state dependency and variability of the
drift coefficient.} The inference performance of SNN using PCM synapses could be further improved by reducing the inherent conductance drift from the
devices. The recently demonstrated projected-PCM cell architecture with an order of magnitude lower drift coefficient is a promising step in this
direction\cite{Y2015koelmansNatComm, Y2018giannopoulosIEDM}.

\section*{Discussion}
One of the key questions that we have evaluated in this work is the ability of stochastic analog memory devices to represent the synaptic strength in
SNNs that have been trained to create spikes at precise time instances. As opposed to supervised learning in second generation ANNs whose network
output is determined typically by normalization functions such as softmax, learning to generate multiple spikes at precise time instants is a harder
problem. Compared to classification problems, the accuracy of which depends only the relative magnitude of the response from one out of several
output neurons, the task here is to generate close to 1000 spikes at the desired time instances over a  period of 1250\,ms from 168 spiking neurons,
which are only excited by 132 spike streams. Furthermore, the high correlation observed among several  input spike-streams (due to the inherent
correlations present in the frequency components of the input audio signal) also makes the learning problem challenging for networks with
low-precision weights. While the spike rate based pixel intensity plots clearly represents the desired images, we chose to evaluate our training
performance using an accuracy metric defined in terms of spike time tolerance, since SNNs designed to process precise spike times rather than spike
rates could be expected to have higher energy efficiency and smaller response time.

At the same time, the observed conductance characteristics of biological synapses is not all too different from those exhibited by our nanoscale
phase change memory devices. 
 PCM device conductance changes in a stochastic manner when programmed using
partial-SET pulses, and the conductance saturates in approximately $ 16-20 $ pulses corresponding to a bit precision on the order of $ 4-5 $ bits.
Synaptic transmission  in biology is also observed to be stochastic  and quantized, and previous   studies have estimated that biological synapses
have a precision of about $4.6$ bits\cite{Bartol2015}.

However, a major    difference between our experiments and biology is the dynamics of the spiking neurons and the learning algorithms used for weight
updates. We have implemented the highly simplified leaky-integrate-and-fire model with an artificial refractory period to model the neuronal
dynamics. Numerous studies have pointed out that neuronal integration and spiking in biology is a highly non-linear and error-tolerant process, with
the most striking behavior revealed by the experiments of Mainen and Sejnowski showing extremely reliable spiking behavior of neocortical neurons
when excited by noisy input currents\cite{Mainen1503}. Such non-linear behaviors may also play a key role in allowing biological networks to create
spikes with more reliability and precision.

While several algorithms have been developed from mathematical formulations of cost-functions involving spike rates and spike times, the mechanisms
employed by nature to achieve the same task are still not well-understood. Most of the neuroscience literature focuses on local learning rules such
as hebbain plasticity, STDP, triplet-STDP, etc. It is not clear how these different local unsupervised learning rules come together to enable
biological networks to encode and process information using precise spike times. Neverthless, the artificial algorithms being developed are achieving
increasing amounts of success in showing software-equivalent performance in several common benchmark tasks in machine learning.

In summary, we analyzed the potential of the PCM devices to realize synapses in SNNs that can learn to generate spikes at precise time instances via
large scale (approximately 180,000 PCMs) supervised training and inference experiments and simulations. We proposed several strategies to improve the
performance of these PCM based learning networks  to compensate for the device level non-idealities. For example,  synapse update granularity
{improved via} multi-PCM  configurations  { can improve the training accuracy}. Also, the performance drop during inference due to the conductance
drift could be compensated via array level scaling based on a global factor which is a function of the time elapsed since training alone. We
successfully demonstrate that in spite of its state-dependent conductance update and drift behavior, PCM synaptronic networks could be trained to
generate spikes with a few milliseconds of precision in SNNs. In conclusion, PCM based computational memory presents a promising candidate to realize
energy efficient bio-mimetic parallel architectures for processing time encoded SNNs in real time.

\section*{Methods}
\subsection*{Audio to spike conversion}
The silicon cochlea chip has 64 band-pass filters with frequency bands logarithmically distributed from 8\,Hz to 20\,kHz and generates spikes
representing left and right channels. Further, due to the synchronous delta modulation scheme used to create the spikes, there were on-spikes and
off-spikes. The silicon cochlea generated spikes with a time resolution of 1\,$ \mu $s. The spikes were further sub-sampled to a time resolution of
0.1\,ms. The final input spike streams used for the training experiments have an average spike rate of 10\,Hz. Combining all the filter responses
with non-zero spikes for left and right channels and the on and off spikes, there are 132 input spike streams.

\subsection*{Neuron model}
The SNN output neurons were modeled using leaky-integrate and fire (LIF) model. Its membrane potential $ V(t) $ is given by the differential equation
\begin{equation*}
C_m\frac{dV(t)}{dt} = -g_L(V(t)-E_L)+ I(t)
\end{equation*}
where $ C_m $ is the membrane capacitance, $ g_L $ is the leak conductance, $ E_L $ is the leak reversal potential, and $ I(t) $ is the net synaptic
current flowing into the neuron. When $ V(t) $ exceed a threshold voltage $ V_T $, $ V(t) $ is reset to $ E_L $ and a spike is assumed to be
generated. Once a spike is generated, the neuron is prevented from creating another spike within a short time period called refractory period $
t_{ref} $. For the training experiment,  we used $ C_m = 300\, $pF, $ g_L = 30\, $nS, $ E_L = -70\,$mV, $ V_T = 20\,$mV, $ t_{ref} = 2\,$ms. For the
NormAD training algorithm, the approximate impulse response of the LIF neuron is given as $ \frac{1}{C_m}e^{-t/\tau_L} u(t) $ where $ \tau_L =
0.1\times C_m/g_L$ and $ u(t) $ is the Heaviside step function. During training, the neuron responses were simulated using $ 0.1\, $ms time
resolution.

\subsection*{PCM platform}
The experimental platform is built around a prototype chip of 3 million PCM cells. The PCM devices are based on doped-Ge$ _2 $Sb$ _2 $Te$ _5 $
integrated in 90\,nm CMOS technology \cite{Y2007breitwischVLSI}. The fabricated  PCM cell area is 50\,F$ ^2 $ (F is the feature size for the 90\,nm
technology node), and each memory device is connected to two parallel 240\,nm wide n-type FETs. The chip has circuitry for cell addressing, ADC for
readout, and circuits for voltage or current mode programming.

The PCM chip is interfaced with the Matlab workstation via FPGA boards and a high-performance analog-front-end (AFE) board. AFE implements digital to
analog converters, electronics for power supplies, and voltage and current references. FPGA board implements digital logic for interfacing PCM and
AFE board and perform data acquisition. A second FPGA board has an embedded processor and Ethernet unit for overall system control and data
management.

\subsection*{Experiment}
The SNN training problem was initially simulated using double-precision (FP64) synapses in the Matlab simulation environment. The weight range for
the SNN was approximately in the range [-6000, 6000]. To map the weights to the PCM conductance values in a multi-PCM configuration, the conductance
range contribution from each device is assumed to be [0.1\,$ \mu $S, 8\,$ \mu $S]. The conductance values are read from the hardware using a constant
read voltage of 0.3\,V, scaled to the network weights and are used for matrix-vector multiplications in the software simulator. When different number
of PCM device are used per synapse, a scaling factor is determined such that the total conductance map to the same weight range. The weight updates
determined from the training algorithm at the end of an epoch is programmed to the PCM devices using  partial-SET pulses of a duration 50\,ns and
amplitudes in the range [40\,$ \mu $A, 130\,$ \mu $A]. The device conductance values are read after each epoch and is used to update the SNN synapse
values. Since each conductance values are read and programmed in series, each training epoch was emulated in an average of \unit[6.3]{s}.

For inference, the PCM conductance values were read at logarithmic time intervals after 100 epochs of training and the effect of compensation schemes
were evaluated in the software simulator.

\vspace{1cm}

%

\section*{Acknowledgments}
We would like to thank Dr. Shih-Chii Liu  from the Institute of Neuroinformatics, University of Zurich, for technical assistance with converting the
audio input to spike streams using a silicon cochlea chip. A.S. acknowledges support from the European Research Council through the European Unions
Horizon 2020 Research and Innovation Program under grant number 682675. B.R. was supported partially by the National Science Foundation through the grant 1710009 and Semiconductor Research Foundation through the grant 2717.001.

\section*{Author Contributions}
B.R. and A.S. conceived the main ideas in the project. S.R.N and A.S. designed the experiment and S.R.N performed the simulations. I.B. and S.R.N
performed the experiment. M.L.G and I.B. provided critical insights. S.R.N and B.R. co-wrote the manuscript with inputs from other authors. B.R.,
A.S., and E.E. directed the work.

\end{document}